\documentclass[fleqn,usenatbib]{mnras}

\usepackage{newtxtext,newtxmath}

\usepackage[T1]{fontenc}

\DeclareRobustCommand{\VAN}[3]{#2}
\let\VANthebibliography\thebibliography
\def\thebibliography{\DeclareRobustCommand{\VAN}[3]{##3}\VANthebibliography}

\usepackage{graphicx}	
\usepackage{amsmath}	

\title[The nuclear environment of NGC 2442]{The nuclear environment of NGC 2442: a Compton-thick low-luminosity AGN}

\author[Patr\'icia da Silva et al.]{
Patr\'icia da Silva,$^{1}$\thanks{\href{mailto:p.silva2201@gmail.com}{p.silva2201@gmail.com}}
R. B. Menezes$^{2}$\thanks{\href{mailto:roberto.menezes@maua.br}{roberto.menezes@maua.br}}
Y. Díaz$^{3}$\thanks{\href{mailto:yaherlyn.diaz@postgrado.uv.cl}{yaherlyn.diaz@postgrado.uv.cl}}
Elena L\'opez--Navas,$^{3}$
J. E. Steiner$^{1}$
\\
$^1$Instituto de Astronomia, Geof\'isica e Ci\^encias Atmosf\'ericas, Departamento de Astronomia, Universidade de S\~ao Paulo, 05508-090, SP, Brazil\\
$^2$Instituto Mau\'a de Tecnologia, Pra\c{c}a Mau\'a 1, 09580-900, S\~ao Caetano do Sul, SP, Brazil \\
$^{3}$Instituto de F\'isica y Astronom\'ia, Facultad de Ciencias, Universidad de Valparaíso, Gran Bretaña No. 1111, Playa Ancha, Valparaíso, Chile}

\date{Accepted: 2021 April 27. Received: 2020 December 29}

\pubyear{2021}

\begin{document}
\label{firstpage}
\pagerange{\pageref{firstpage}--\pageref{lastpage}}
\maketitle

\begin{abstract}

The detailed study of nuclear regions of galaxies is important because it can help understanding the active galactic nucleus (AGN) feedback mechanisms, the connections between the nuclei and their host galaxies, and ultimately the galaxy formation processes. We present the analysis of an optical data cube of the central region of the galaxy NGC 2442, obtained with the integral field unit (IFU) of the Gemini Multi-Object Spectrograph (GMOS). We also performed a multiwavelength analysis, with \textit{Chandra} data, \textit{XMM--Newton} and \textit{NuSTAR} spectra, and \textit{Hubble Space Telescope} (\textit{HST}) images. The analysis revealed that the nuclear emission is consistent with a Low Ionization Nuclear Emission-line Region (LINER) associated with a highly obscured compact hard X-ray source, indicating a Compton-thick AGN. The \textit{HST} image in the F658N filter (H$\alpha$) reveals an arched structure corresponding to the walls of the ionization cone of the AGN. The gas kinematic pattern and the high gas velocity dispersion values in the same region of the ionization cone suggest an outflow emission. The stellar archaeology results indicate the presence of only old stellar populations ($\sim$ 10 Gyr), with high metallicity (z = 0.02 and 0.05), and the absence of recent star formation in the central region of NGC 2442, which is possibly a consequence of the AGN feedback, associated with the detected outflow, shutting off star formation. NGC 2442 is a late-type galaxy similar to the Milky Way, and comparisons show that the main difference between them is the presence of a low-luminosity AGN.

\end{abstract}




\section{Introduction}

 The nuclear region of galaxies is one of the most relevant morphological components to be studied and their detailed study can show their connections with the surrounding regions, and better reveal their role in the formation and evolution of galaxies. Such connections are probably directly related to the feeding and feedback mechanisms from active galactic nuclei (AGNs; see \citealt{thaisa1,thaisa2} and references therein). These connections were made evident from the correlations between, for example, the masses of the central supermassive black holes (SMBHs) and parameters of the host galaxies, such as the stellar velocity dispersion of the bulge (which results in the \textit{M--}$\sigma$ relation; \citealt{fer00,geb00,gul09}). Outflows from AGNs represent a manifestation of the feedback mechanism and, in many situations, can shut off star formation \citep{fab12}. However, under specific circumstances, outflows can actually do the opposite and induce star formation \citep{mai17,gal19}. The characterization of the stellar populations in the regions around AGNs, therefore, has a significant importance for the study of the effects of outflows. The feeding mechanism of AGNs, on the other hand, may be shown by the presence of circumnuclear features, such as nuclear spirals and nuclear rings (e.g. \citealt{paper613}).

 This work presents the analysis of NGC 2442 nucleus (focusing on the inner central $\sim$ 400 pc). This source is a SAB(s)bc galaxy \citep{RC3} and it is located at 21.8 $\pm$ 1.4 Mpc \citep{cartier2017}; considering this value for the distance of the galaxy, 1 arcsec $\sim$ 106 pc. Besides being a Milky Way morphological twin, from which we can delineate some comparison with our Galaxy, NGC 2442 is a great example, worth of being studied, of the interaction between nucleus and the host galaxy. Structures in the nuclear region were detected suggesting the influence of the host galaxy: a ring/disc-like structure was detected from the analysis of the CO rotation curve with a radius of about 12.5 arcsec (1.3 kpc, considering the galaxy distance) by \citet{bajaja99}. Besides that, \citet{pancoast10} claimed that the nuclear region of this galaxy is the most obscured of the galaxy and that the morphology of the H$\alpha$ emission resembles a spiral structure that is located, mostly, inside the circumnuclear ring, with radius of about 0.8 kpc, observed in 8 $\mu$m. From its nuclear activity, NGC 2442 was classified as a Low Ionization Nuclear Emission-Line Region (LINER) from optical emission-line ratios by \citet{bajaja99}.

\begin{figure*}
\begin{center}

  \includegraphics[scale=0.27]{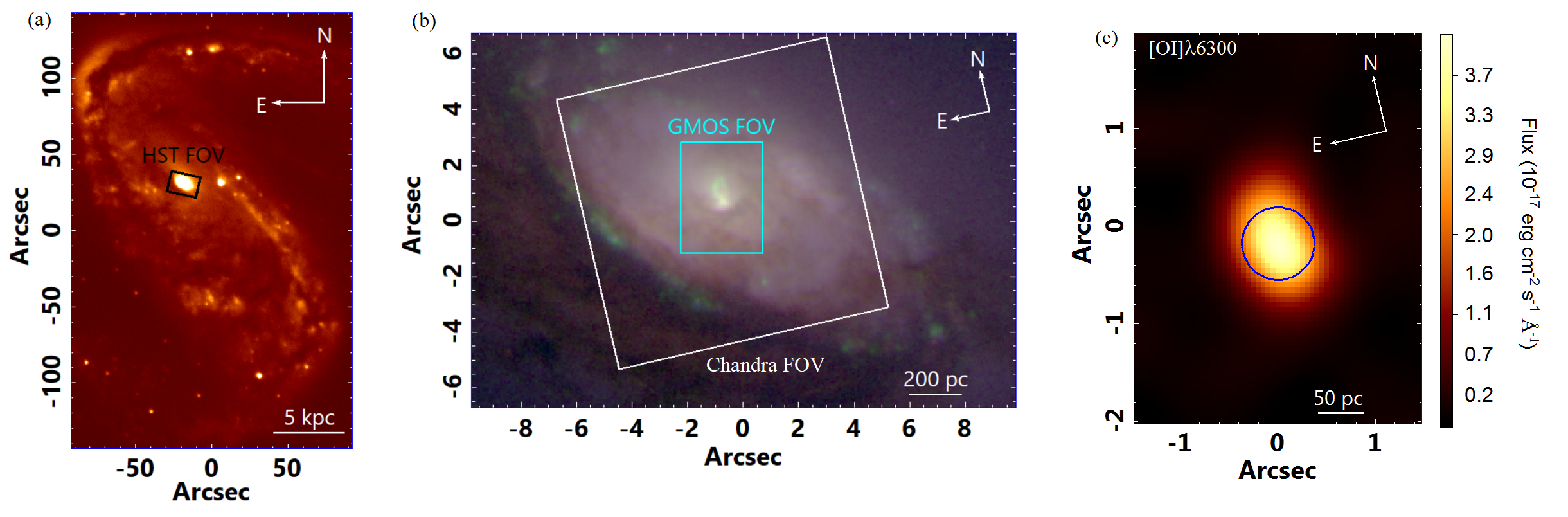}
  \caption{Panel (a): image of NGC 2442 obtained with the Cerro Tololo Inter-American Observatory's 1.5-meter telescope at the \textit{B} band. The black rectangle represents the FOV of the \textit{HST} images. Panel (b): RGB composition with the filters F435W in blue, F658N in green, and F814W in red from \textit{HST}. The white square represents the FOV of \textit{Chandra} data and the cyan rectangle the FOV of the GMOS data cube. The total size of this image is 13.5 $\times$ 1.6 arcsec$^2$ (with a scale of $\sim$ 1400 pc). Panel (c): image of the integrated flux of [O~\textsc{i}]$\lambda$6300 emission line from the GMOS data cube, indicating the nucleus (blue circle- whose size is equal to the PSF of the observation). The area of GMOS FOV is 2.95 $\times$ 4 arcsec$^2$ (with a scale of $\sim$ 400 pc). The position angle (PA) of the GMOS observation is 167$\degr$ and the \textit{HST} images were oriented with the same PA, while the \textit{Chandra} data have PA = 0$\degr$.  \label{NGC2442_ctio}}
  
\end{center}
\end{figure*}

We present here a more detailed analysis of the central region of NGC 2442, in order to examine the interactions between the nucleus and the host galaxy and also to analyse the nature of the nuclear emission of this galaxy. We perform a multiwavelength study with the use of optical data, with a high spatial resolution data cube obtained with the Gemini Multi-Object Spectrograph (GMOS) from the Gemini-South telescope and images from the \textit{Hubble Space Telescope} (\textit{HST}), and X-ray data, with a \textit{Chandra} data cube and spectra obtained with the Nuclear Spectroscopic Telescope Array (\textit{NuSTAR}; \citealt{Harrison2013}) and the X-ray Multi-Mirror Mission (\textit{XMM--Newton}). With the use of these data, we are able to analyse the coexistence of the various structures observed in this nuclear region. This work is organized as follows: section~\ref{sec_obs} describes the observations and data treatment; section~\ref{sec_emissionlineratio} shows the results obtained from the analysis of the optical emission-line ratios and of the X-ray emission in the central region of NGC 2442; section~\ref{secHST} is focused on the \textit{HST} data; section~\ref{sec_cinematicagas} shows the analysis of the gas kinematics of the GMOS data cube; section~\ref{sec_starlightresults} presents the results of the spectral synthesis used to study the stellar archaeology; in section~\ref{sec_discussion}, the results are discussed; and section~\ref{sec_conclusions} presents the main conclusions of this work. Finally, Appendix~\ref{mapas_starlight} shows further results from the spectral synthesis applied to the GMOS data cube, and Appendix~\ref{modelsxray} presents the details of the models used in the analysis of the X-ray nuclear spectra of this galaxy.

\section{Observations and data reduction}\label{sec_obs}

The main data for this study were obtained with the IFU of the GMOS on Gemini-South telescope, in the one-slit mode. These data are part of the Milky Way morphological twins sample of the \textit{Deep IFS View of Nuclei of Galaxies} (DIVING$^{3D}$) survey, which has the goal of analysing the nuclei of all southern-hemisphere galaxies with $B < 12$ and $|b|>15\degr$ (\citealt{divingproceedings} and Steiner, J. E. et al., in preparation).  

These data were taken on 2014 February 23 as part of the observation program GS-2014A-Q-5. Three $\sim$ 816~s exposures were taken with spatial dithering (with dither steps of 0.2 arcsec) and PA = 167$\degr$. The R831+G5322 grating was used, centred at 5850~\AA, resulting in a spectral resolution of R = 4340 and in a spectral coverage from 4785 to 6821~\AA. The reduced data cube spatial pixels (spaxels) size is 0.05 arcsec.

The data reduction was performed using the Gemini package in \textsc{iraf} environment. The reduction processes are listed in \citet{rob3}. After the creation of three data cubes, we applied the treatment described by \citet{rob1}, and \citet{rob2,rob3}, which consists of the following procedures: correction of the differential atmospheric refraction, combination of the three data cubes in a median in order to have one, Butterworth spatial filtering \citep{gwoods}, instrumental fingerprint removal and Richardson--Lucy deconvolution \citep{rich,lucy}. The full width at half-maximum (FWHM) of the point spread function (PSF) used in the deconvolution process was estimated from the acquisition image observed at 6300~\AA\ and its value is 0.74 arcsec. All those treatment procedures were performed using Interactive Data Language (\textsc{idl}) scripts developed by our group.

Other complementary data, obtained from public archives, were used, such as the \textit{Chandra X-ray Observatory} data (0.2 -- 10.0 keV), \textit{HST} images, and \textit{XMM--Newton} (0.25 -- 10.0 keV) and \textit{NuSTAR} (3.0 -- 79.0 keV) spectra. The purpose of including these observations is to improve the study of the nucleus of the galaxy with a multiwavelength analysis.

The \textit{Chandra} data were taken, using the ACIS-I instrument \citep{acis_referencia}, on 2015 July 3 with exposure time of 39.56 ks, under the observation program number 15610062 [principal investigator (PI): Garmire, G.]. A data cube was created from the events table using an \textsc{idl} script, with spaxels with sizes of 0.492 arcsec, and we performed a spatial re-sampling, in order to obtain spaxels with sizes of 0.246 arcsec. After that, a Butterworth spatial filtering was applied, since the spatial re-sampling can introduce high spatial frequency noise. 

The observations with \textit{XMM--Newton} of NGC 2442 were taken in 2017 in three different epochs. Due to the low count rates of the source, we chose the longest observation for the analysis, with an exposure time of 60ks (observation ID: 0794580101). The observation data files (ODFs) from the European Photon Imaging Camera (EPIC) on the detector were processed using the Science Analysis System (\textsc{sas}--version 17.0.0). We followed standard procedures to obtain calibrated and concatenated event lists, by filtering them for periods of high background flaring activity and by extracting light curves and spectra. The source events were extracted using a circular region with a radius of 49 arcsec centred on the target, and the background events were extracted from a circular region with a radius of 98 arcsec on the same chip far from the source. The data were taken using the small window mode and the medium blocking filter to reduce pile-up and optical loading effects, respectively. We verified that the photon pile-up is negligible in the filtered event list with the task \textsc{epatplot}. After that the response matrix files (RMFs) and ancillary response files (ARFs) were generated and the spectra were re-binned in order to include a minimum of 25 counts in each background-subtracted spectral channel and in order to not oversample the intrinsic energy resolution by a factor larger than 3.

Since NGC 2442 has three observations with \emph{NuSTAR}, we selected the one with the highest  exposure time (80~ks -- observation ID: 80202029004), taking into account the low count rate in this galaxy. The data were processed using \textsc{nustardas} v1.6.0, available in the  \emph{NuSTAR} Data Analysis Software. The event data files were calibrated with the \textsc{nupipeline} task using the response files from the Calibration Data Base \textsc{caldb} v.20180409 and \textsc{heasoft} version 6.25. With the \textsc{nuproducts} script, we generated both the source and background spectra, plus the ARF and RMF files. For both focal plane modules (FPMA and FPMB), we used a circular extraction region with a radius of 50 arcsec centred on the position of the source. The background selection was made taking a region free of sources of twice the radius of the target. Spectral channels were grouped with the \textsc{ftools}  task \rm{grppha} to have a minimum of 20 counts per spectral bin. The source is significantly detected in the 3 - 60~keV energy range.

The \textit{HST} images were taken on 2006 October 29 using the advanced camera for surveys (ACS) under the program 10803 (PI: Smartt, S.). The exposure time for each filter was different: 1580 s for the F435W filter, 1350~s for the F658N filter, and 1200~s for the F814W filter. The size of the image spaxels is 0.05 arcsec. In order to compare these data with the GMOS data cube images, we convolved the F435W filter image with the PSF of the GMOS data cube. After that, we cut the F435W filter image in a way that the nucleus (the position of the stellar emission peak) was at the same position as in the image of the blue continuum (4785\AA\ -- 4924\AA) of the GMOS data cube and their fields of view (FOVs) were the same. This cut was then applied to the other filter images as well.

Figure \ref{NGC2442_ctio}, panel (a) shows the whole galaxy with the indication of the \textit{HST} FOV and panel (b) shows the RGB composition with the three filters of the \textit{HST}, where we can see a nuclear ring and the morphology of the nuclear region. In this image we also indicated the FOV of the \textit{Chandra} and GMOS data. Lastly, panel (c) shows the main emitting region (represented by the image of [O~\textsc{i}]$\lambda$6300). The blue circle represents the nucleus of NGC 2442 galaxy and the area from which we extracted the spectrum to study the nature of its emission.

\subsection{Spectral synthesis and the gas data cube}\label{sec_starlight}

In order to study the stellar archaeology, we applied a spectral synthesis to the GMOS data cube with the \textsc{starlight} software \citep{starlight}. The stellar population spectral base used for this procedure was derived from the Medium-resolution Isaac Newton Telescope Library of Empirical Spectra -- MILES \citep{blazquez}, called CB2007. We also used the results from the spectral synthesis performed with \textsc{starlight} in order to create a synthetic stellar data cube. This data cube was subtracted from the original one to obtain a data cube with mainly gas emission, which we call gas data cube.  

\begin{figure}
\begin{center}

  \includegraphics[scale=0.43]{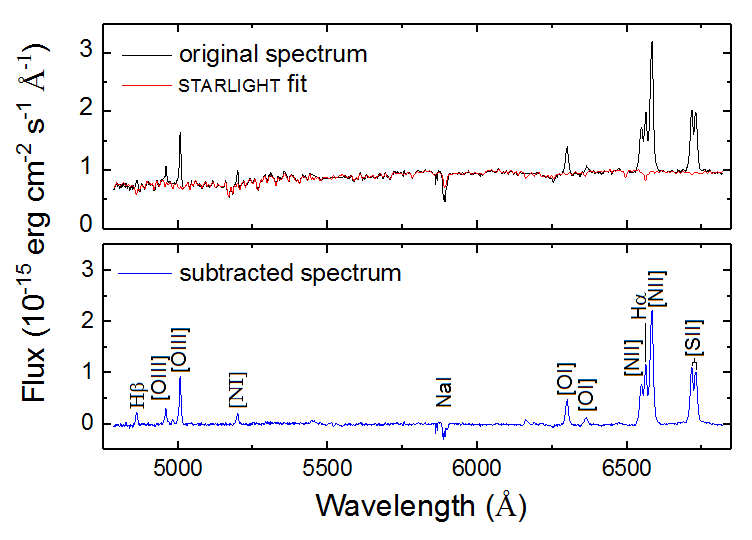}
  \caption{Spectrum of the nucleus of the galaxy  (identified in the image of [O~\textsc{i}]$\lambda$6300 in  Fig.~\ref{NGC2442_ctio}c) of the original data cube after the treatment and the spectral synthesis fit (performed with the \textsc{starlight} software), which is shown in red. The spectrum of the nucleus after the stellar emission subtraction is shown in blue with the main emission lines in the optical band identified.  \label{NGC2442_espectro_regiao1total}}
  
\end{center}
\end{figure}

Fig.~\ref{NGC2442_espectro_regiao1total} shows the spectrum of the nucleus of NGC 2442 (see Fig.~\ref{NGC2442_ctio}c), the \textsc{starlight} fit of this spectrum and the resulting emission-line spectrum obtained by subtracting the fit from the original spectrum.

\section{ The nuclear region: a Compton-thick low-luminosity AGN}\label{sec_emissionlineratio}

In order to establish the nature of emission in the central region of NGC 2442, we analyse the optical and the X-ray emission in the nuclear and circumnuclear regions of this galaxy.

\subsection{Optical emission-line ratios}

From the GMOS gas data cube it is possible to study the spatial morphology of the optical line emission. The main emitting region in this data cube is the nucleus and it comprises the emission of [O~\textsc{i}]$\lambda$6300, an emission-line typical of partially ionized regions that are mainly located near AGNs. We extracted the spectrum of a circular region (represented by the blue circle in Fig.~\ref{NGC2442_ctio}c), centred at the peak of the [O~\textsc{i}]$\lambda$6300 emission with a radius equal to half of the FWHM of the PSF of the observation in order to avoid the contamination of the circumnuclear emission. The spectrum of the nucleus is shown in  Fig.~\ref{NGC2442_espectro_regiao1total}, in which we can see that some emission lines are blended, and the blending of the [N~\textsc{ii}]+H~$\alpha$ emission lines might be caused by a broad component of the H~$\alpha$ emission line. Such broad component, however, is not seen in the H~$\beta$ emission line.

\begin{figure*}
\begin{center}

  \includegraphics[scale=0.3]{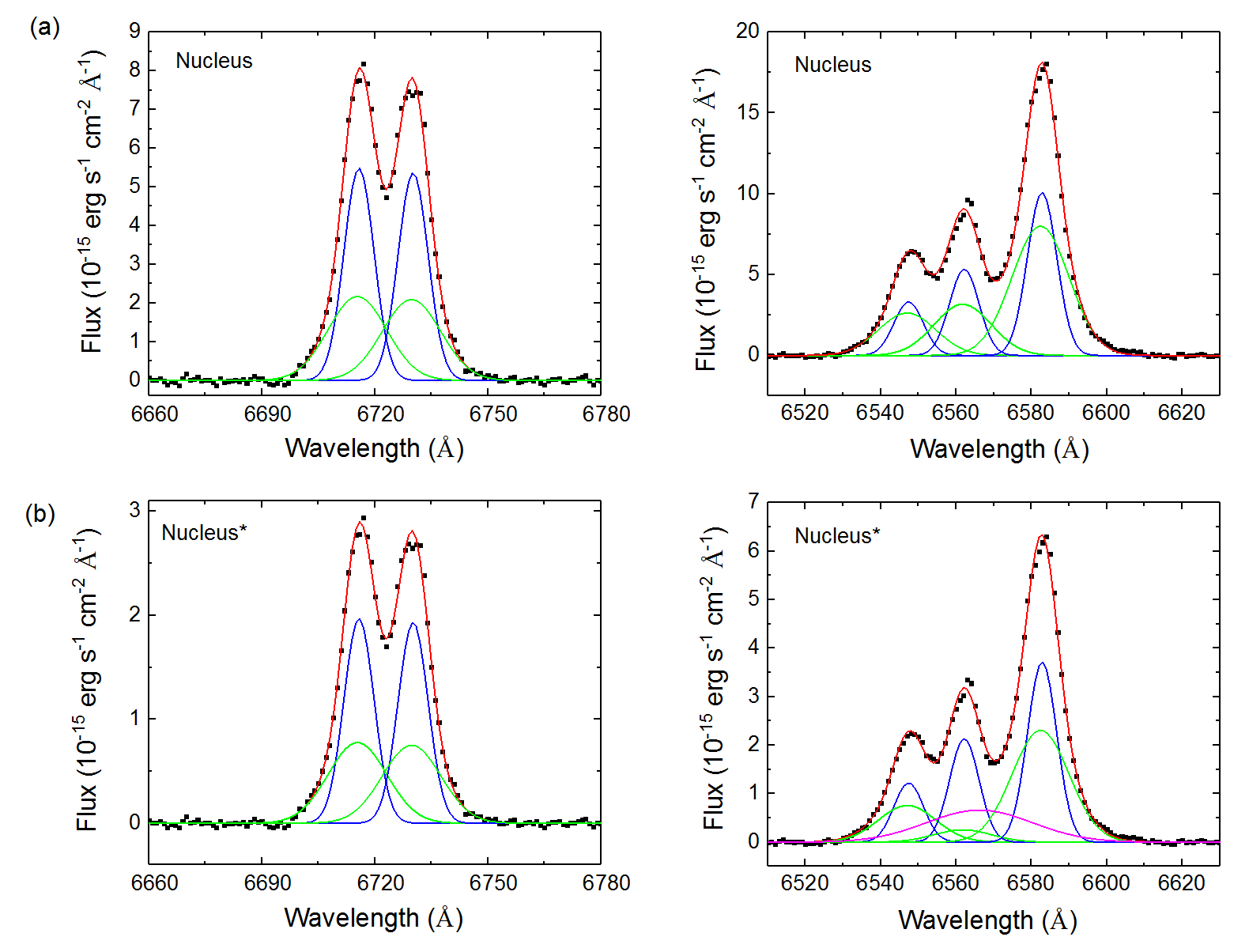}
  \caption{Gaussian decomposition of the spectrum of the nucleus without a broad component of the H$\alpha$ emission line (panel a) and with this broad component (panel b). Each emission line was decomposed as a set of two Gaussian functions (the green and blue curves) and the magenta curve represents the broad component. The results indicated with "*" take into account the broad component of the H$\alpha$ emission line. In order to calculate the integrated flux of the emission lines, the integrated flux of the blue + green curves of each line was taken. \label{decomposicaoregiao1_2442}}
  
\end{center}
\end{figure*}

In order to calculate emission-line ratios and perform a diagnostic diagram analysis \citep{diagramabpt,veilleux} of the nuclear spectrum, the integrated fluxes of the non-blended emission lines were obtained by direct integration. On the other hand, the integrated fluxes of the blended emission lines [N~\textsc{ii}]$\lambda\lambda$6548,6583, H$\alpha$ and [S~\textsc{ii}]$\lambda\lambda$6716,6731 were determined by fitting each emission line with a sum of two Gaussian functions. First, we fitted two sets of two Gaussian functions for the [S~\textsc{ii}]$\lambda\lambda$6716,6731 doublet (a set is composed by a blue and a green Gaussian -- see Fig.~\ref{decomposicaoregiao1_2442}). Then we used the results of the parameters of those lines (redshift and width of each component -- green and blue) to fit the others ([N~\textsc{ii}]+H$\alpha$). The spectrum was corrected for extinction, by using the H$\alpha$/H$\beta$ emission-line ratio (assuming the intrinsic value for H$\alpha$/H$\beta$ = 2.86) and the extinction law of \citet{cardelli}. 

We opted to perform the Gaussian decomposition of the [N~\textsc{ii}]+H$\alpha$ lines with and without including a broad Gaussian (representing the broad component of H$\alpha$). Both results were quite acceptable, since their reduced $\chi^{2}$ values were 2.8 and 3.6, respectively; therefore, we are presenting both. Fig.~\ref{decomposicaoregiao1_2442} shows the final results (after the extinction correction) of these two decompositions.

\begin{figure}
\begin{center}

  \includegraphics[scale=0.4]{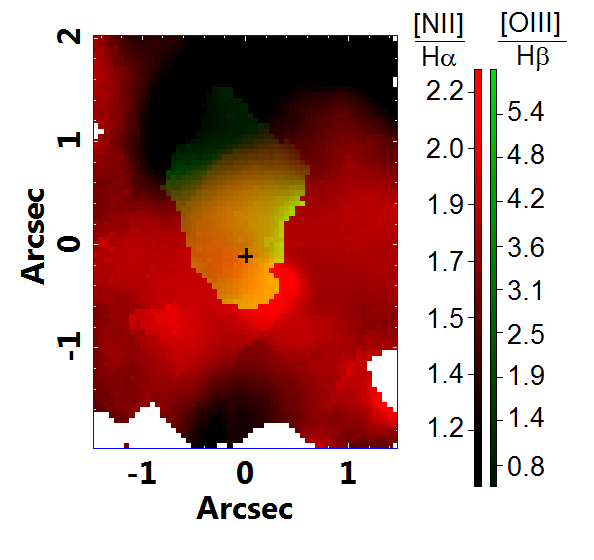}
  \caption{RG composition showing the maps of the emission line ratios of [N~\textsc{ii}]$\lambda$6583/H$\alpha$ in red and [O~\textsc{iii}]$\lambda$5007/H$\beta$ in green. The areas where the A/N ratio of any of the emission lines used to calculate the ratios is lower than 5 were masked.\label{line_ratios_maps}}
  
\end{center}
\end{figure}

Table~\ref{tablerazao2442} presents the emission-line ratios of the nucleus and shows that, regardless of the broad component of the H$\alpha$ emission line, the nuclear line emission of NGC 2442 is compatible with that of LINERs. This result is consistent with the one obtained by \citet{bajaja99}. The question that remains is if this LINER emission is from gas photoionized by a low-luminosity AGN \citep{fer83,hal83} or by post-AGB stars that can emulate this emission \citep{sta08,era10,cid11}.

In order to analyse the circumnuclear emission in NGC 2442, we calculated the integrated fluxes of the H$\beta$, [O~\textsc{iii}]$\lambda$5007, H$\alpha$ and [N~\textsc{ii}]$\lambda$6583 emission lines and also the [O~\textsc{iii}]$\lambda$5007/H$\beta$ and [N~\textsc{ii}]$\lambda$6583/H$\alpha$ emission-line ratios of each spectrum of the GMOS/IFU data cube. The integrated fluxes of H$\beta$ and [O~\textsc{iii}]$\lambda$5007 were obtained via direct integration. However, the integrated fluxes of H$\alpha$ and [N~\textsc{ii}]$\lambda$6583 were determined by fitting the [N~\textsc{ii}]+H$\alpha$ emission lines with a sum of three Gaussian functions (one Gaussian for each emission line). We used this simpler procedure, instead of that applied to the nuclear spectrum, in order to avoid degeneracies and problems of convergence when fitting the lines of all spectra in the data cube. The spectra were also corrected for the interstellar extinction, using the H$\alpha$/H$\beta$ values and the extinction law of \citet{cardelli}.

Fig.~\ref{line_ratios_maps} shows an RG composition of the maps of the [O~\textsc{iii}]$\lambda$5007/H$\beta$ and [N~\textsc{ii}]$\lambda$6583/H$\alpha$ ratios. The values obtained taking into account emission lines with amplitude/noise ratios (A/N) lower than 5 were masked. A diagnostic diagram analysis \citep{bal81,vei87,kew06,sch07} shows that the [N~\textsc{ii}]/H$\alpha$ values do not indicate the presence of H~\textsc{ii} regions around the central AGN. Meanwhile, although a large portion of the [O~\textsc{iii}]$\lambda$5007/H$\beta$ map was masked (due to the low A/N ratio of the H$\beta$ in most of the FOV), we can see that there are considerably high values of this ratio on a specific region that includes the nucleus and the ionization cone (see section \ref{secHST}). The emission-line ratios in the non-masked regions are all compatible with those of LINERs in both maps.

\begin{table}
\begin{center}

\caption{Emission-line ratios and the H$\alpha$ luminosity of the nucleus of NGC 2442 from the GMOS data cube. }
\label{tablerazao2442}
\begin{tabular}{ccc}

\hline
Emission-line ratios                                  & Nucleus      & Nucleus*        \\ \hline
{[}OIII{]}$\lambda$5007/H$\beta$                               & 2.81 $\pm$ 0.12 & 2.98 $\pm$ 0.13                       \\ 
{[}NII{]}$\lambda$6583/H$\alpha$                               & 2.23 $\pm$ 0.10 & 3.1 $\pm$ 0.3       \\ 
{[}OI{]}$\lambda$6300/H$\alpha$                                & 0.460 $\pm$ 0.019 & 0.67 $\pm$ 0.7       \\ 
({[}SII{]}$\lambda$6716+$\lambda$6731)/H$\alpha$  & 1.71 $\pm$ 0.08 & 2.7 $\pm$ 0.3        \\ 
{[}SII{]}$\lambda$6716/{[}SII{]}$\lambda$6731     & 1.03 $\pm$ 0.05 &  1.03 $\pm$ 0.04      \\ 
 H$\alpha$ luminosity (10$^4$ $L_{\bigodot}$) & 167 $\pm$ 7 &  38 $\pm$ 4          \\ \hline
\end{tabular}
\end{center}
\textit{Notes.}{ \footnotesize The spectrum was corrected for extinction by using the H$\alpha$/H$\beta$ ratio.The two results shown were obtained with (*) and without a broad H$\alpha$ component in the fit (we did not detect broad emission in the H$\beta$ emission line).}

\end{table}

\subsection{X-ray emission} \label{sec_raiosX}

\begin{figure*}
\begin{center}

  \includegraphics[scale=0.42]{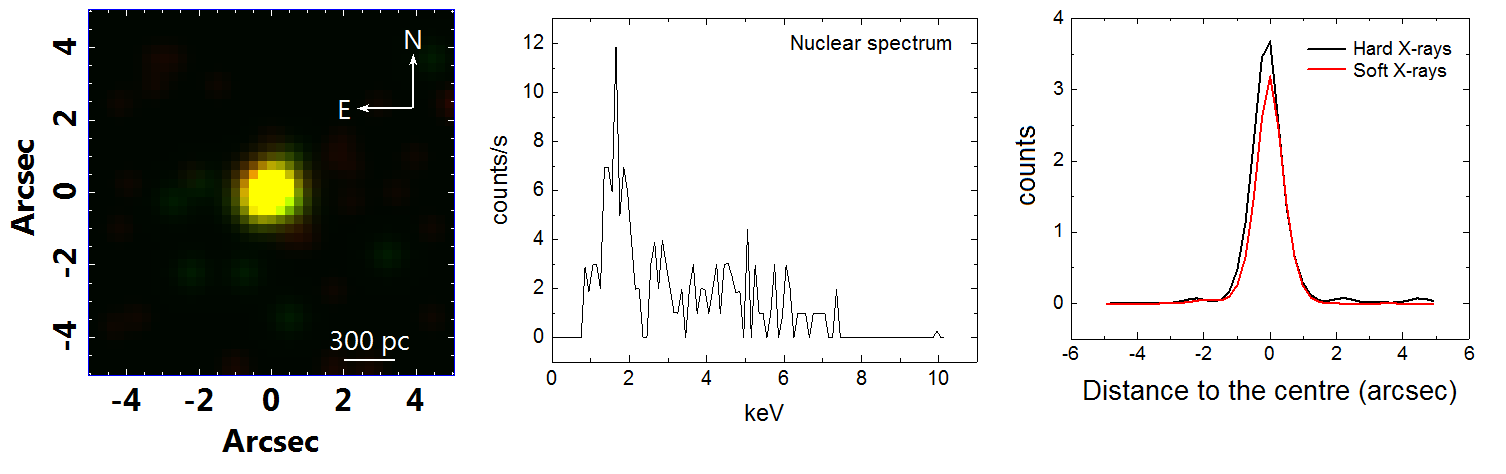}
  \caption{RG composition of the images of hard X-rays (in red) and soft X-rays (in green). The spectrum of the nuclear region (extracted from a circular area with radius of 6 pixels, or 1.5 arcsec, larger than the PSF of these data estimated from the emission of Mrk 202 -- see \citealt{paper613}) shows a considerable emission of soft X-rays. The comparison between the horizontal profiles of both emissions show that there is an excess of hard X-rays emission at the centre and both emissions come from the same source, which is the AGN. The centre was defined as the peak of emission of hard X-rays.\label{xrays2442}}
\end{center}
\end{figure*}

In order to analyse the high-energy end of the nuclear emission of NGC 2442, we created a data cube from the data retrieved from the \textit{Chandra X-ray Observatory} archive (for more details see section \ref{sec_obs}). An RG composite image, representing the soft and hard X-ray emission, the nuclear spectrum and brightness profiles of this data cube are shown in Fig.~\ref{xrays2442}. 

By analysing the RG composition of soft (0.5 -- 2 keV) and hard (2 -- 10 keV) X-rays images we see that both emissions come from a compact source at the nucleus and that the hard X-ray emission is very strong, which indicates that the LINER emission that we observe in the optical data is actually from a low-luminosity AGN. 

The low signal-to-noise ratio (S/N) in the hard X-ray portion of the spectrum in Fig.~\ref{xrays2442} suggests a significant obscuration of the AGN. Such obscuration, in the X-rays band, would also cause the decrease of emission of soft X-rays. However, when we look at the spectrum of the nuclear region, we see that, comparatively, the emission in soft X-rays is more intense than that in hard X-rays, although the emission in hard X-rays seems to be more intense at the centre (when we compare the horizontal profiles). Such results might indicate that the AGN emission is considerably obscured (possibly Compton-thick) by the dust torus, or/and by the gas along the line of sight (as the \textit{HST} images indicate). On the other hand, the ionization cone (see Figs.~\ref{NGC2442_HST}b and ~\ref{NGC2442_HST_imenoshalpha}b) is not being obscured and it is the main source of the soft X-ray emission. The low spatial resolution does not allow, in this case, to separate the ionization cone and nucleus, as we see in the \textit{HST} images (see section~\ref{secHST}).

Since the \textit{Chandra} data analysis was not conclusive about the AGN obscuration, we tried to investigate the X-ray emission of this source further with \emph{XMM–Newton} and \emph{NuSTAR} spectra. The aim of this analysis is to put constrains on the column density in this source, in order to explore the Compton thickness. We followed the method used by \citet{dia20}, where the reflection is analysed by fitting different reflection models. The spectral analysis of the data was performed using \textsc{xspec} version 12.10.0 \citep{arnaud96}. A visual inspection showed low values of counts at energies above 50.0 keV, which were rejected from the data. All the errors correspond to 90$\%$ unless otherwise noted. The details about the model used here are described in Appendix~\ref{modelsxray}.

\begin{figure}
\begin{center}

  \includegraphics[scale=0.35]{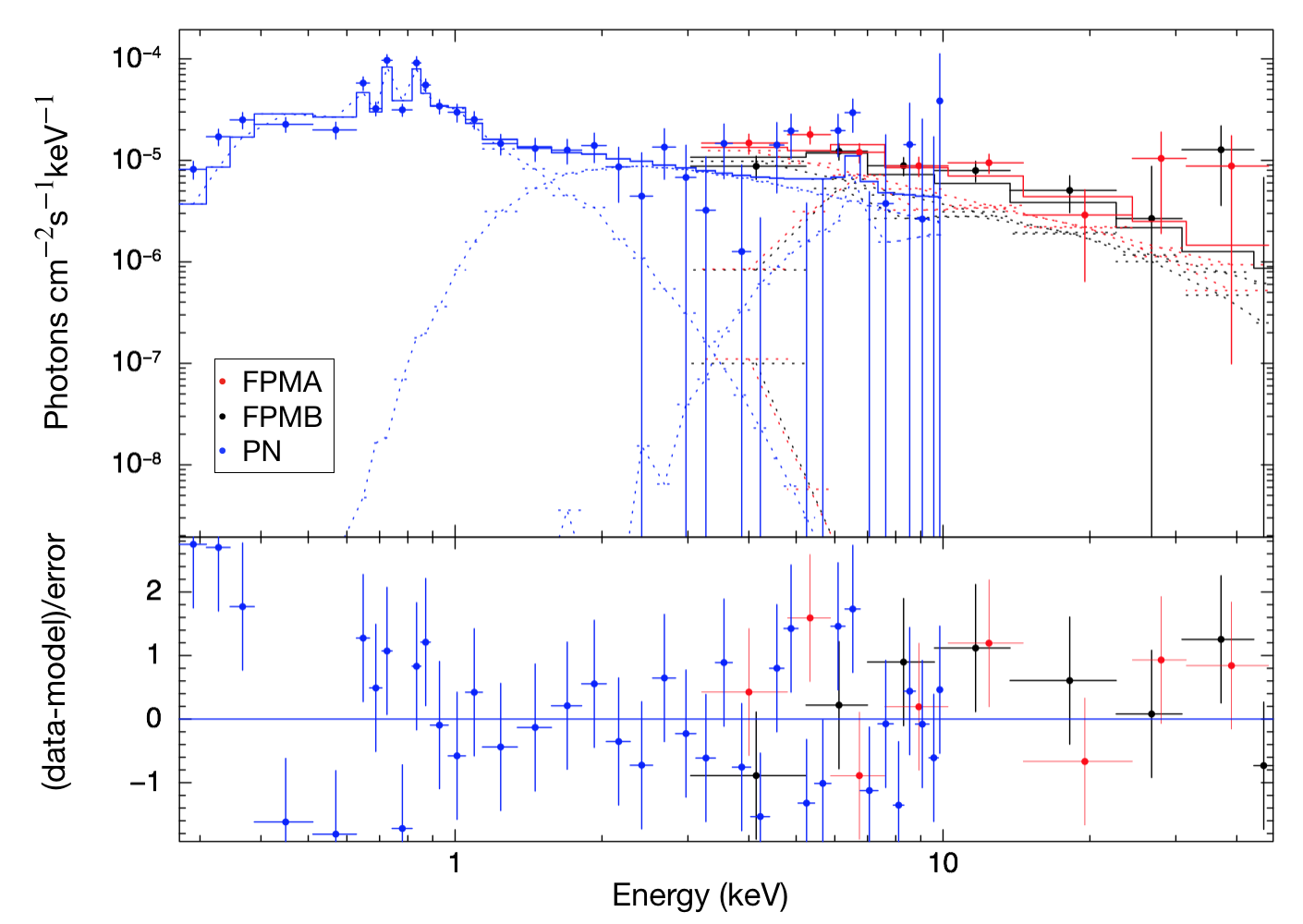}
  \caption{Upper panel: Best-fitting Borus+cut-off PL model (solid line) to the \emph{NuSTAR} FPMA and FPMB focal planes and \textit{XMM--Newton} Pn spectra of NGC 2442 (filled circles). Lower panel: residuals in terms of (data-model)/error. \label{spec_borus2442}}
  
\end{center}
\end{figure}

\begin{figure}
\begin{center}

  \includegraphics[scale=0.3]{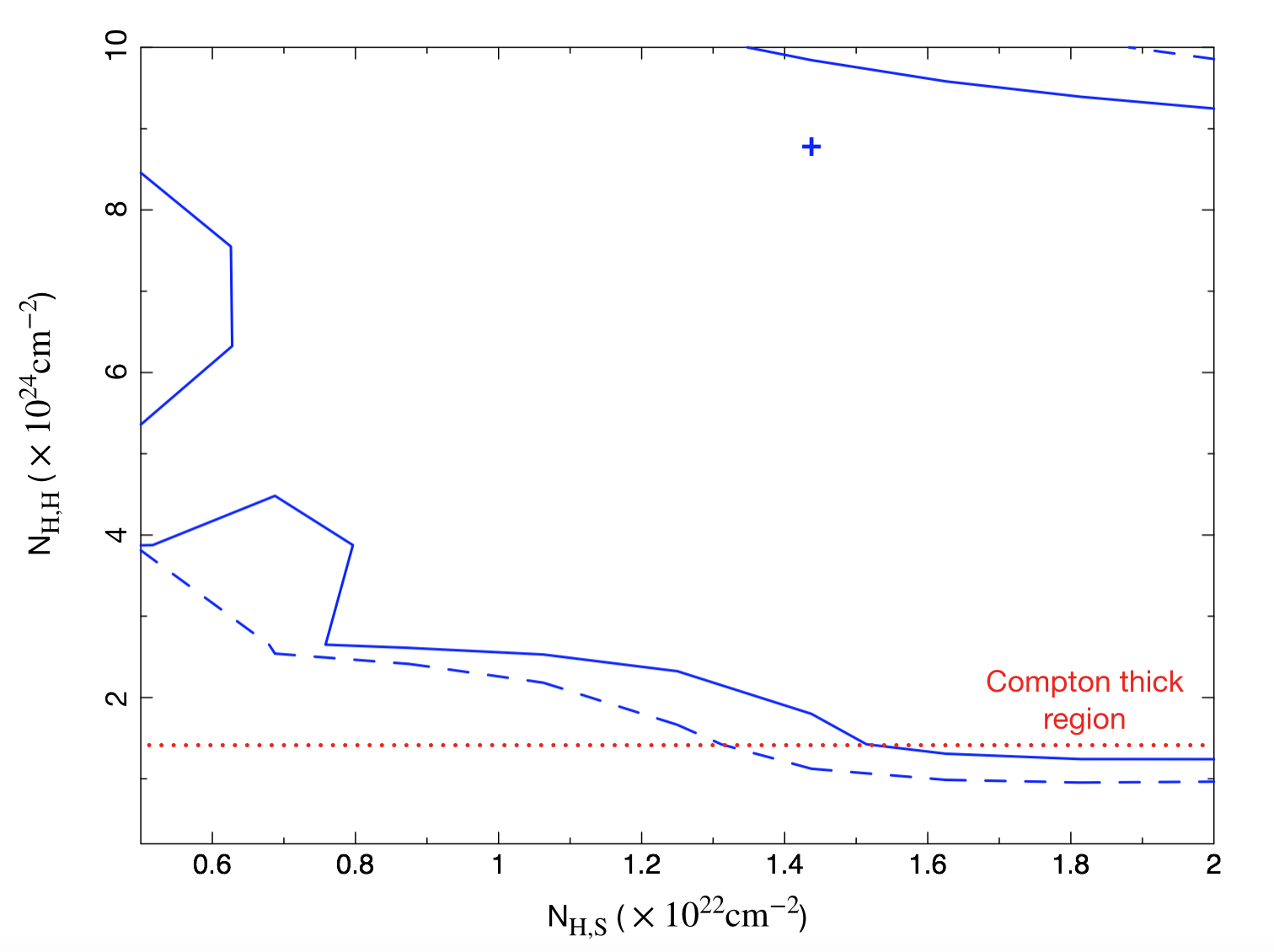}
  \caption{Two-dimensional $\Delta\chi^{2}$ contours for the column density in the scatter and nuclear component in NGC 2442. The blue line represents the 68$\%$ interval confidence and the 90$\%$ is shown with the blue dashed line. The ``+'' symbol represents the best-fitting value, and the red pointed line delimits the Compton-thick and the Compton-thin regions. \label{nh2442}}
  
\end{center}
\end{figure}

The best-fitting model is shown in Fig. \ref{spec_borus2442}.  This fit is statistically acceptable with $\chi^2$=203.25 for 205 d.o.f. and has a $\Gamma$=2.19$_{-0.19}^{+0.23}$ and E$_{\rm{cut}}$ <2000 keV. The column density of the absorber acting on the extended scattered power law in this fit is $N_{\rm{H,S}}$=1.41$_{-0.55}^{+1.12}$ $\times$10$^{22}$ cm$^{-2}$ and in the nuclear power law $N_{\rm{H,H}}$>2.49 $\times$10$^{24}$ cm$^{-2}$ showing a column density consistent with a Compton-thick source, as can be seen in Fig.~\ref{nh2442}. It is worth mentioning that $N_{\rm{H,H}}$ takes into account the obscuration caused by a an inner torus (the reflector) and also by clouds located at farther distances from the AGN. The calibration uncertainties between the instruments are C$_{\rm{FPMA/FMPB}}$=1.25$_{-0.14}^{+0.19}$ and C$_{\rm{Nu/XMM}}$=0.66$_{-0.22}^{+0.06}$.

We were able to put constraints also on the column density of the reflector and its covering factor: the reflector is constrained to have a column density of $\log(N_{H,tor})$=23.65$^{+0.17}_{-0.53}$  for a covering factor CF=0.31$_{-0.07}^{+0.25}$ with an unconstrained inclination between the observer and the polar axis, showing values below 90\degr with 90$\%$ of confidence.

The previous analysis indicates that this structure is obscured. However, we could not put constrains in the inclination between the torus and the observer, in which case the Compton thickness of the galaxy would be better explained with the high column density in the line of sight ($N_{\rm{H,S}}$ and $N_{\rm{H,H}}$). The previous results point out that the nucleus of NGC 2442 harbours an obscured Compton-thick AGN.

\section{Circumnuclear ring and the ionization cone observed with the \textit{HST}}\label{secHST}

\begin{figure*}
\begin{center}

  \includegraphics[scale=0.45]{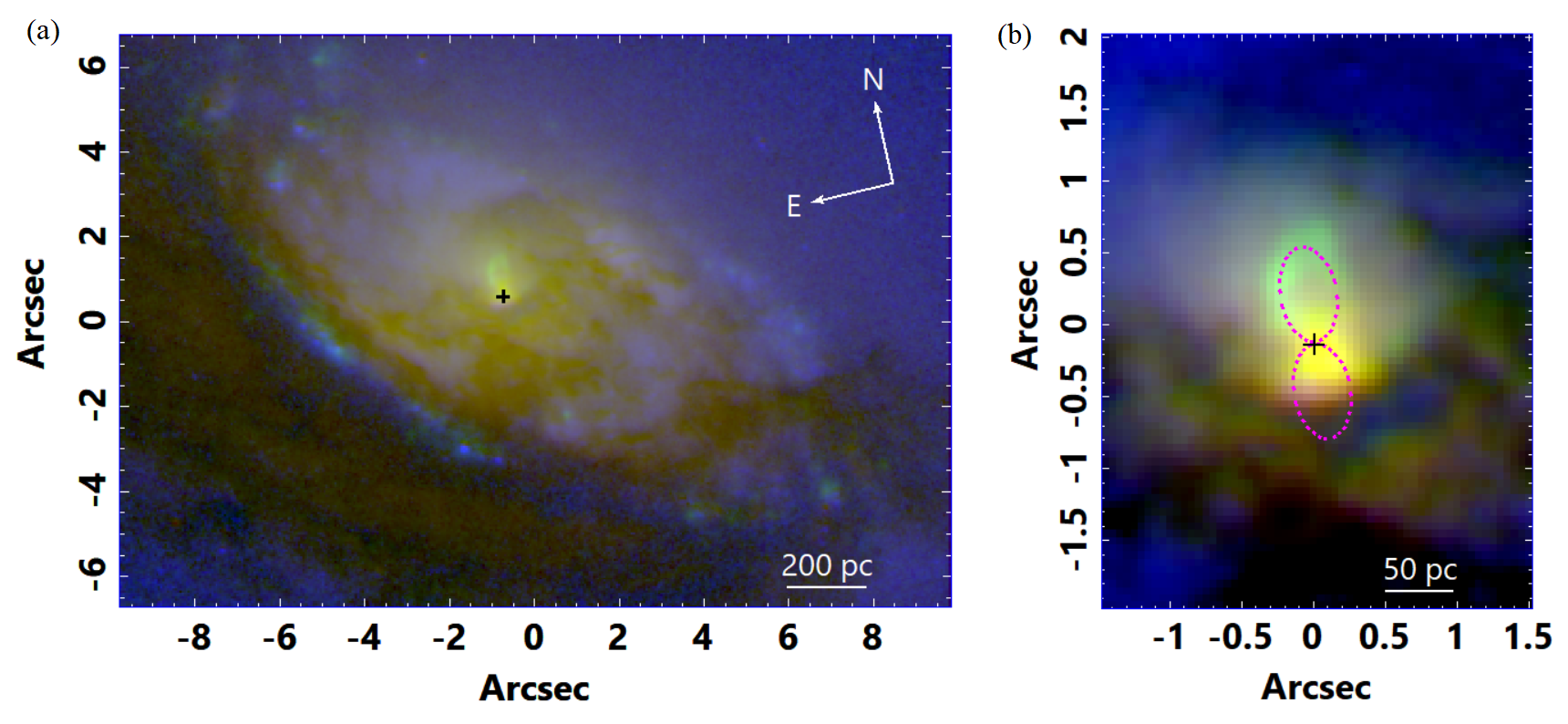}
  \caption{ Panel (a): RGB composition of filter F814W in red, filter F658N in green and F814W - F435W  in magnitude scale (equivalent to $I - B$) in blue. Panel (b): the same RGB composition with the GMOS FOV. The black cross represents the coordinates of the centre of [O~\textsc{i}]$\lambda$6300 emission in the \textit{HST} images (estimated from the superposition of the GMOS and \textit{HST} data). The size of this cross in panel (a) does not represent the uncertainty, only in panel (b) where the pixels are resolved (and its size represents the uncertainty of 3$\sigma$). The magenta dashed line delineates the arched structure of the F658N (H$\alpha$) emission, which is being interpreted as the walls of the ionization cone. The other side of the ionization cone cannot be seen, due to the high extinction towards the south. \label{NGC2442_HST}}
  
\end{center}
\end{figure*}

In order to visualize in more detail the optical emitting structures in a larger FOV and also with better spatial resolution, we analysed \textit{HST} images from the \textit{HST} public archive (for more information, see section \ref{sec_obs}). Fig.~\ref{NGC2442_HST}(a) shows an RGB composite image of the central 13.5 $\times$ 19.6 arcsec$^2$ ($\sim$ 1.4 $\times$ 2 kpc$^2$) of NGC 2442, revealing an apparent stellar/gaseous disc. There is also a lot of obscuration or redder emission towards the east portion of the FOV and a significant redder emission in the disc itself that looks like streams and clouds towards the nucleus. When we look at Fig.~\ref{NGC2442_HST}(b), which is the same image with the GMOS data cube FOV size, we note more clearly the central structure of the H$\alpha$ emission (F658N filter -- in green) that shows a possible gas disc at the bottom and an arched structure that extends towards north of the nucleus. This arched structure is possibly associated with the emission from the walls of the ionization cone of the AGN in the nucleus (for an example of ionization cone walls, see \citealt{may_hourglass}). We can also see that there is a more intense red emission/extinction in the southern region. This extinction or redder emission is associated with this gas disc that seems to have a different inclination when compared with the larger disc structure seen in Fig.~\ref{NGC2442_HST}(a).

\begin{figure*}
\begin{center}

  \includegraphics[scale=0.37]{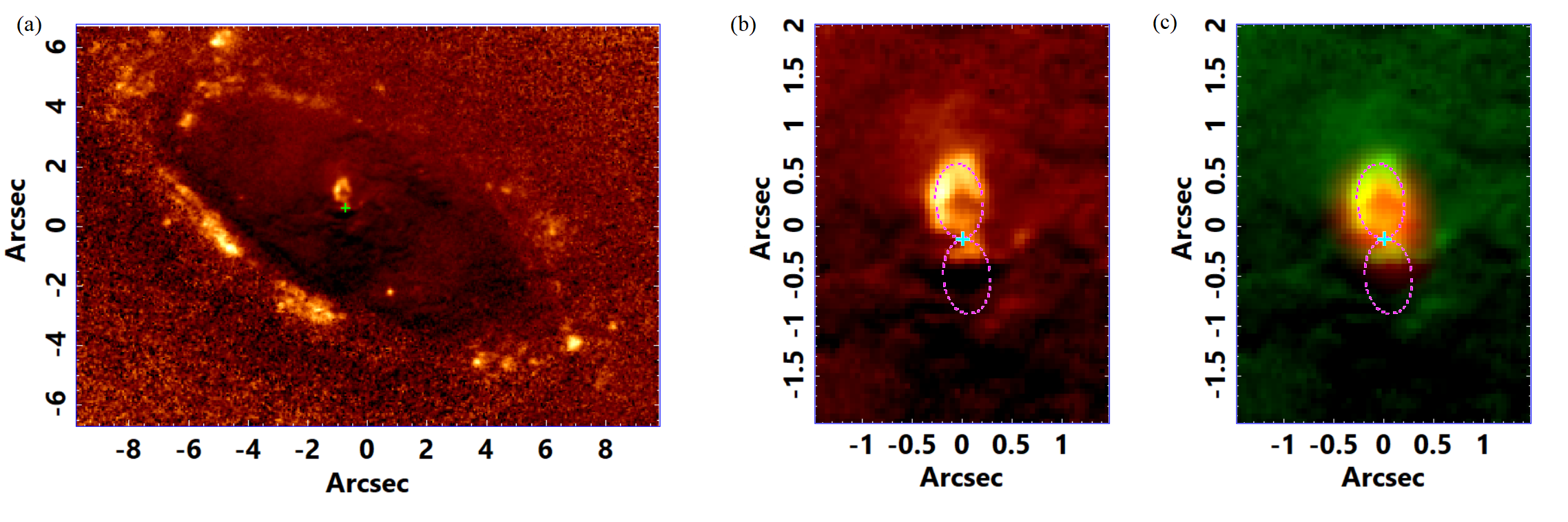}
  \caption{Image of F814W - F658N in magnitude scale (equivalent to I-H$\alpha$). The clear areas represent where the emission of H$\alpha$ is more significant (panel~a) in the \textit{HST} FOV considered here and (panel b) in the GMOS data cube FOV. As in Fig.~\ref{NGC2442_HST}(b), in panel (b), there is the indication of the hourglass structure representing the ionization cones. Panel (c): RGB composition of the image of [O~\textsc{iii}]$\lambda$5007 in red and the image of panel (b) in green. The position of the centre of [O~\textsc{i}]$\lambda$6300 is indicated by the green and cyan crosses, whose size represents the 3$\sigma$ uncertainty.\label{NGC2442_HST_imenoshalpha}}
  
\end{center}
\end{figure*}

Fig.~\ref{NGC2442_HST_imenoshalpha}(a), which represents essentially the (inverted) continuum-subtracted H$\alpha$ emission (filters F814W - F658N), shows that there is a very clumpy ring structure (already detected in previous studies -- \citealt{bajaja99,pancoast10}). Such a ring has an apparent elliptical morphology, with an estimated major axis radius of 8.7~$\pm$~0.5 arcsec (979~$\pm$~80~pc) with a PA of 44$\degr$~$\pm$~4\degr and a minor axis radius of 3.51~$\pm$~0.24 arcsec (371~$\pm$~35~pc). Such an estimate was made by plotting some ellipses in the \textit{HST} image of this ring (Fig.~\ref{NGC2442_HST_imenoshalpha}a) and its centre is compatible with the position of the [O~\textsc{i}]$\lambda$6300 emission peak. If the ring is circular its radius is equal to 8.7~$\pm$~0.5 arcsec (979~$\pm$~80~pc) with an inclination of 66\degr\!\!.2~$\pm$~2\degr\!\!.3. 

When we look at Fig.~\ref{NGC2442_HST_imenoshalpha}(c), we note again that the most intense emission of [O~\textsc{iii}]$\lambda$5007 comes from the same region where the arched structure detected in H$\alpha$ is located (Fig.~\ref{NGC2442_HST_imenoshalpha}b). We can also see the same structure on the map of [O~\textsc{iii}]$\lambda$5007/H$\beta$ (Fig.~\ref{line_ratios_maps} in green). This supports the hypothesis that such an arched structure corresponds to the walls of the ionization cone (represented by the emission of [O~\textsc{iii}]$\lambda$5007). There is a high extinction towards the south, which is probably responsible for the obscuration of the other side of the ionization cone and the AGN. Based on the size and morphology of the arched structure, we delineated an hourglass scheme to represent the ionization cone and, as mentioned before, we can only see one side of it (the northern one). 

\section{Gas kinematics: a rotating disc and outflows} \label{sec_cinematicagas}

\begin{figure*}
\begin{center}

  \includegraphics[scale=0.4]{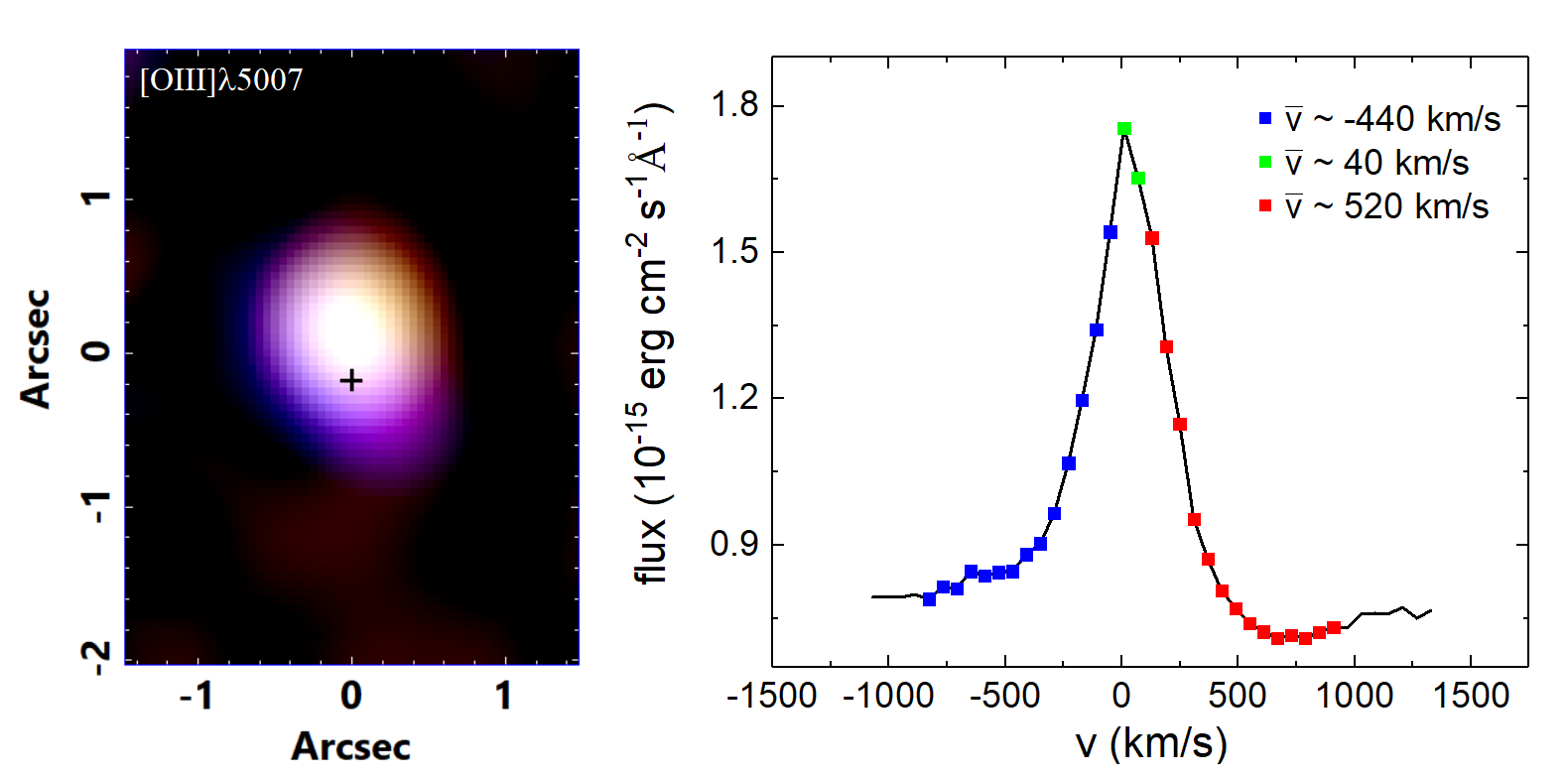}
  \caption{RGB composition of the [O~\textsc{iii}]$\lambda$5007 emission line from the GMOS data cube. The line profile extracted from the spectrum of the nucleus is shown with the points of the wavelength interval of each colour of the RGB (and each mean velocity). The black cross represents the position of the centre of [O~\textsc{i}]$\lambda$6300 and its size the 3$\sigma$ level uncertainty. One can note that the green colour image has the morphology and size of the PSF of the data. This is because such emitting region is very compact and probably associated with the arched structure (the ionization cone wall), which, by its turn, cannot be resolved with the GMOS spatial resolution. Since the green colour represents low velocity emission, this region has low velocities, compatible with the scenario that such arched emission is associated with the ionization cone walls. \label{NGC2442_gmos_kinematics}}
  
\end{center}
\end{figure*}

By studying the emission lines of the GMOS/IFU data cube, it is possible to analyse the gas kinematics, which is very important to evaluate possible feedback effects of the AGN in circumnuclear regions. One way of gathering such information is by studying the line profiles. The RGB composition of [O~\textsc{iii}]$\lambda$5007 of Fig.~\ref{NGC2442_gmos_kinematics} represents quite well the main emission lines kinematic behaviour, and shows that the region with low-velocity gas (in green) is nearly coincident with the region associated with the walls of the ionization cone seen in the \textit{HST} images. The areas with gas with redshifted (in red) and blueshifted (in blue) emission are opposite to each other, but there is also a superposition of both emissions at the centre, and the region with gas with redshifted emission is also coincident with the areas corresponding to the ionization cone, more than the blueshift emission. The wavelength intervals for the redshift and blueshift images taken from the line profile have the same length and the moduli of their velocities are relatively close to each other. 

\begin{figure*}
\begin{center}

  \includegraphics[scale=0.27]{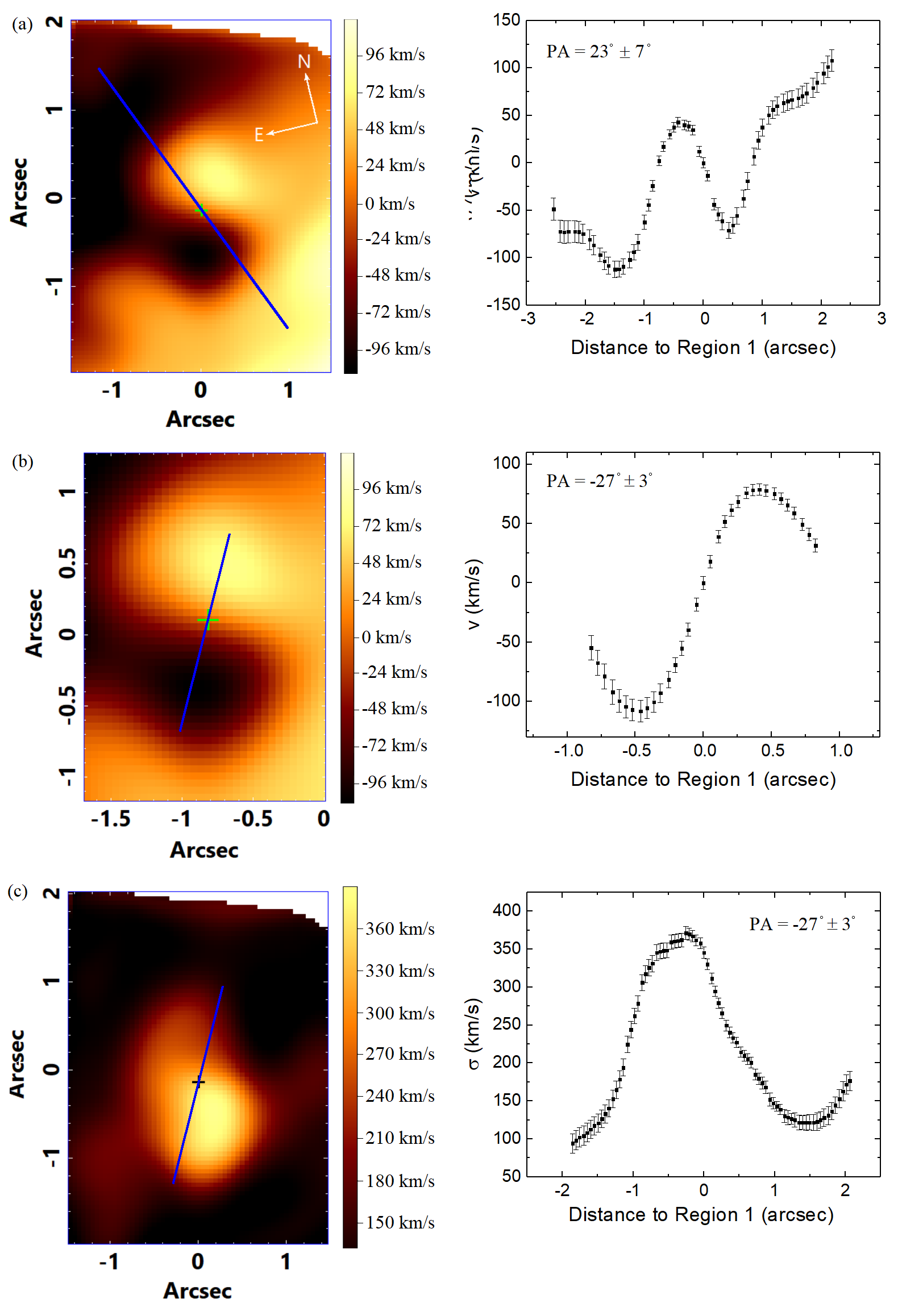}
  \caption{Velocity maps obtained from the emission lines H$\alpha$+[N~\textsc{ii}]$\lambda$6548, 6583 of the whole GMOS FOV (panel a) and of the inner centre (panel b), with their respective curves extracted from different PAs, since the direction of the phenomena seems to differ in those two regions. The map and curve of the gas velocity dispersion is shown in panel (c). The curve of the velocity dispersion values was taken from the same PA as in panel (b). The green and black crosses represent the centre of [O~\textsc{i}]$\lambda$6300 emission, which is assumed to have v = 0 km s$^{-1}$, and its size represents the 3$\sigma$ uncertainty. The blue lines in the maps on the left correspond to the axes whose PAs are shown on the right. Such axes connect the minimum and maximum values of the two kinematic phenomena detected in the velocity maps. The uncertainties of the PAs of these axes were determined from the uncertainties of points in the maps with the minimum and maximum values. The white regions in the maps have A/N ratios of the [N~\textsc{ii}]$\lambda$6548 emission line lower than 3 and were not considered in the analysis.\label{NGC2442_gmos_HANII_kinematics}}
  
\end{center}
\end{figure*}

Another way of studying the gas kinematics is  through velocity maps of the emission lines. In this case we fitted the [N~\textsc{ii}]$+$H$\alpha$ emission lines and obtained the maps of Fig.~\ref{NGC2442_gmos_HANII_kinematics}. Such lines were fitted with a sum of three Gaussian functions (each line was fitted with one Gaussian), with the same radial velocity and width. Since this fit was applied to the spectrum corresponding to each spaxel of the gas data cube, the results were the maps of the gas radial velocity and of the gas velocity dispersion. Considering that the procedure was applied taking into account three emission lines simultaneously, we obtained reliable values for the kinematic parameters for all the spectra with A/N of the [N~\textsc{ii}]$\lambda$6548 emission line higher than 3. The uncertainties were estimated using a Monte Carlo procedure. For each spectrum, a histogram of the spectral noise, based on a wavelength interval without emission lines, was constructed. Then, a Gaussian function was fitted to the histogram. Finally, Gaussian distributions (with the same width of the Gaussian fitted to the histogram) of random noise were created and added to the original spectrum and the Gaussian fitting procedure was applied to each obtained “noisy” spectrum. This resulted in a number of gas radial velocity and gas velocity dispersion values. The uncertainties of these kinematic parameters were taken as the standard deviations of all the obtained values.

Fig.~\ref{NGC2442_gmos_HANII_kinematics}(a) shows the velocity map of the [N~\textsc{ii}]$+$H$\alpha$ emission lines and its associated curve extracted from the line of nodes of the map. We can see a pattern consistent with a gas rotation in a large scale together with another kinematic phenomenon in the nuclear region, where the curve and map become irregular. 

When we look at Fig.~\ref{NGC2442_gmos_HANII_kinematics}(b), which shows the same map only in the central region with its associated curve, we can see that there is a bipolarity of velocities that could also be a gas rotation with a different orientation or an outflow. This pattern is similar to the blueshift and redshift features shown by the RGB of [O~\textsc{iii}]$\lambda$5007. The redshift portion is coincident with the region of the ionization cone and the blueshift portion is possibly coincident with the region corresponding to the other side of the ionization cone, which we cannot see, since it is obscured. One can also note that the maximum and the lowest values of the curve are not symmetric, which can indicate a more complex kinematic phenomenon. 

The peak of the gas velocity dispersion is displaced to the south (Fig.~\ref{NGC2442_gmos_HANII_kinematics}c). We can see high velocity dispersion values in the region of the ionization cone as well. This supports the hypothesis of the presence of an outflow in those regions, but could also indicate a discrepancy of the velocities of two rotating discs that are superposed along the line of sight. Given the fact that the location of this peak of dispersion is coincident with the ionization cone region, it is more likely that this represents outflows in the ionization cone.

\section{Stellar archaeology} \label{sec_starlightresults}

\begin{figure}
\begin{center}

  \includegraphics[scale=0.25]{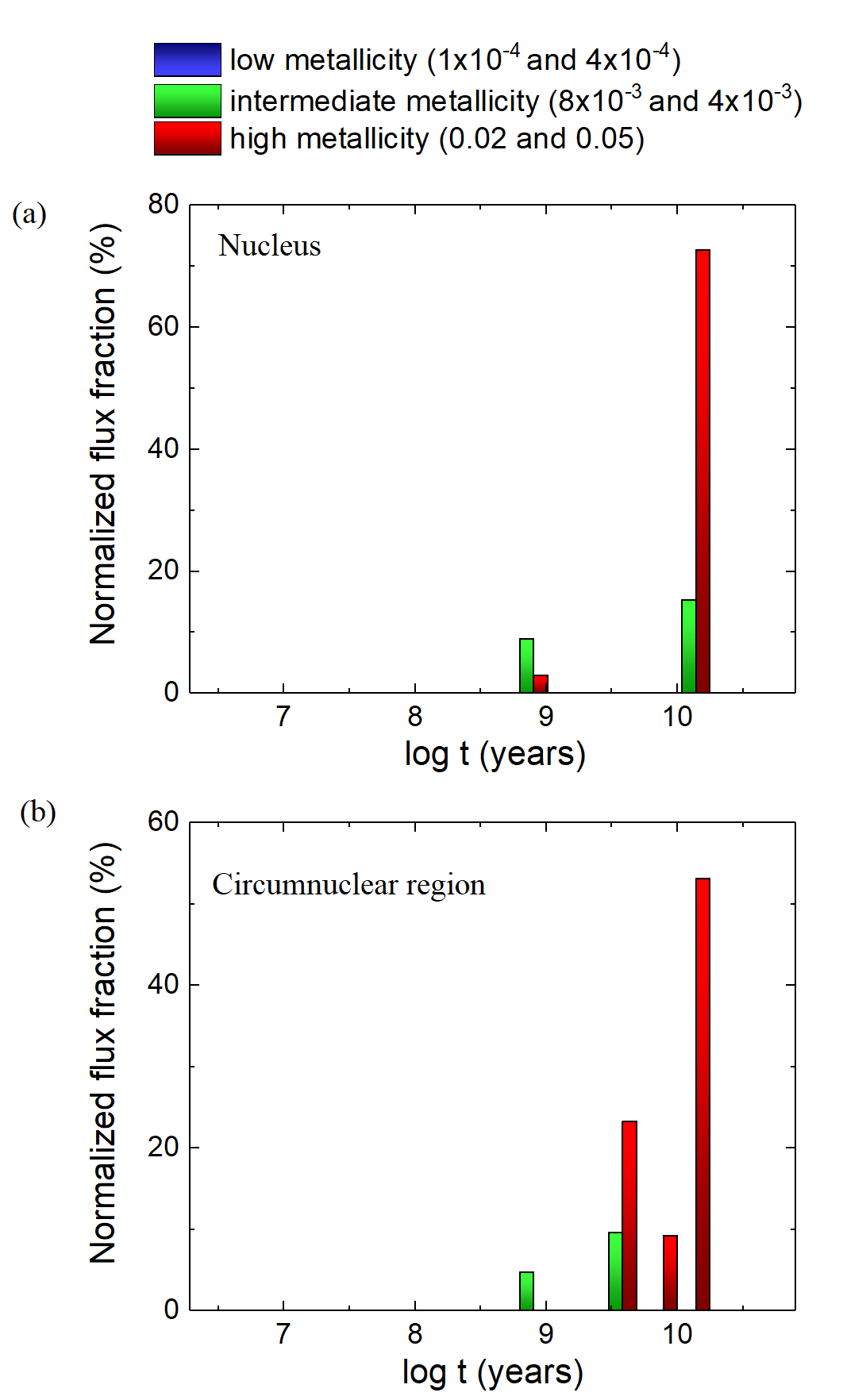}
  \caption{Results of the spectral synthesis applied with the \textsc{starlight} software to (panel a) the spectrum of the nucleus and to (panel b) the spectrum of the circumnuclear region of the GMOS data cube. \label{NGC2442_STARLIGHT}}
  
\end{center}
\end{figure}

In order to study the stellar populations in the central region of NGC 2442, to evaluate possible AGN interference in the star formation in the surrounding regions (in the scale of $\sim$~400~pc), the spectral synthesis was applied with the \textsc{starlight} software (see section \ref{sec_starlight}) to the spectra of two regions: the nucleus and the circumnuclear region, i.e. the result of the subtraction of the spectrum of nucleus from the spectrum of the whole data cube. The histograms in Fig.~\ref{NGC2442_STARLIGHT} summarize the results of this procedure.

The stellar emission of the nucleus of this galaxy (Fig.~\ref{NGC2442_STARLIGHT}a) comes mainly from old ($\sim$~10 Gyr) stars with high metallicity (z = 0.02 and 0.05), representing $\sim~70\%$ of the normalized flux, and intermediate metallicity (z = 0.004 and 0.008), representing $\sim~15\%$ of the normalized flux.

The histogram for the circumnuclear stellar populations does not change much when compared to the one of the nucleus (see Fig.~\ref{NGC2442_STARLIGHT}b). We can see, again, the predominance of old stellar populations with high metallicity, and the main difference is that there is a flux fraction from old stellar populations (between 4 and 8 Gyr, approximately) that are only present in the circumnuclear region.

The method did not find any flux fraction of intermediate age or young stellar populations. Considering that the synthetic stellar spectra provided by the spectral synthesis fitted quite well the observed stellar spectra (see Fig.~\ref{NGC2442_espectro_regiao1total}), we believe that this result is reliable enough to be included in this work and indicates that the nucleus of this galaxy is old. However, this result must be taken with caution, as it is well known that degeneracies can affect the results provided by the spectral synthesis technique. 

In order to have an idea of the uncertainties of the results related to the stellar archaeology provided by the spectral synthesis, we subtracted the resulting synthetic stellar spectra from the observed spectra, which resulted in spectra containing essentially emission lines. After that, for each spectrum, a histogram of the spectral noise was constructed, based on a wavelength interval without emission lines. We fitted a Gaussian function to the histogram and created Gaussian distributions of random noise with the same width of the initial Gaussian fitted. Those distributions of random noise were added to the synthetic spectrum, which resulted in ``noisy'' stellar spectra. The spectral synthesis was applied to each ``noisy'' spectrum and, based on the results, the median age of the detected stellar populations was calculated. For such a calculation, a graph of the cumulative flux fraction as a function of the stellar ages was constructed and the median age was taken as the age of the stellar populations at which the cumulative flux fraction was equal to 50\%. The uncertainty of the ages of the stellar populations was taken as the standard deviation of all median ages obtained from this procedure. The results were 0.10 and 0.15 dex for the nuclear and circumnuclear spectrum, respectively. The same process was applied to estimate the uncertainties of the stellar metallicities and the results were 0.09 and 0.18 dex for the nuclear and circumnuclear spectrum, respectively.

\section{Discussion} \label{sec_discussion}

The analysis of the nuclear region of galaxies has a major importance, as it
reveals connections between the nuclei and the host galaxies, providing
information about their formation and evolution. The nuclear region of NGC 2442 is a very rich environment to be studied and a good example of the influence of the AGN in the host galaxy. As we have seen in the literature \citep{bajaja99} and in this work, this galaxy has a nuclear emission-line spectrum consistent with the presence of a LINER, and we concluded that it is a low-luminosity AGN, since the nucleus of NGC 2442 has a compact hard X-ray source (Fig.~\ref{xrays2442}). The literature also establishes that there is a circumnuclear ring and a possible nuclear spiral in this galaxy \citep{pancoast10}, which are probably associated with feeding mechanisms of the central AGN and, therefore, represent the influence of the host galaxy in the nuclear region.

The analysis of the X-ray data suggests that the central AGN is Compton-thick, although one must be cautious about this result, due to the low number of counts in the data. The modeling of the X-ray spectra did not allow to put constrains on the inclination of the possible inner torus causing most of the obscuration. Even so, considering the significant obscuration detected, we can assume that such a torus is probably nearly edge-on. From the study of the profile of optical emission lines, we cannot determine if this AGN is type 1 or 2, since both emission-line decompositions, with or without a broad-line component, resulted in an acceptable fit. However, according to the unified model \citep{ant93,urr95}, the obscuration of the BLR emission caused by the edge-on torus suggested by analysis of the X-ray data would result in a type 2 AGN. Therefore, the most likely hypothesis, in this case, is that the nucleus of NGC 2442 harbours a type 2 LINER.

The peculiar feature, similar to an arch, detected in the F658N filter image from the \textit{HST} (Fig.~\ref{NGC2442_HST}) corresponds to the walls of the ionization cone of the AGN, considering that it is spatially coincident with the [O~\textsc{iii}]$\lambda$5007 emission (Fig.~\ref{NGC2442_HST_imenoshalpha}) and associated with low-velocity gas (Fig.~\ref{NGC2442_gmos_kinematics}), as revealed by the analysis of the GMOS/IFU data cube. This structure was already seen in other objects, for example in NGC 1068 (see \citealt{may_hourglass}). According to the unified model, the ionizing radiation originating such an ionization cone is collimated by the edge-on torus described before. Therefore, since the axis of the observed ionization cone extends along the north-south direction, this torus must extend along the east-west direction. We suggest that we detect only one ionization cone because the other one (at the opposite side) is highly obscured, as we see in the \textit{HST} images (F814W - F435W, equivalent to $I-B$, in Fig.~\ref{NGC2442_HST_imenoshalpha}). In this case, the obscuring material is much more extended than the edge-on torus and resembles a disc structure around the nucleus. Since this apparent disc also extends along the east-west direction, a plausible hypothesis is that it is an extension of the inner edge-on torus. It is worth mentioning, however, that dusty tori around AGNs are usually not aligned with the discs of the host galaxies (e.g. \citealt{com19}). The dusty extended disc detected in the \textit{HST} images is probably associated with the possible nuclear spiral observed by \citet{pancoast10} (although these two structures have different spatial scales) and, as mentioned before, represents a feeding mechanism of the AGN. Another feeding evidence comes from  the $A_V$ map provided by the spectral synthesis (Fig.~\ref{NGC2442_maps_starlight2}), which shows a spiral pattern towards the AGN. Such a spiral is also possibly connected with the nuclear spiral detected by \citet{pancoast10}, although (similarly to the obscuring disc described above) these structures have different spatial scales.

Besides an apparent large-scale gas rotating disc, the gas velocity map also revealed a bipolar pattern in the region of the arched structure. This, together with the high gas velocity dispersion values detected in the same region (Fig.~\ref{NGC2442_gmos_HANII_kinematics}), suggest the presence of an outflow, which is consistent with the fact that the ionization cone is located in the same area. A natural hypothesis then is that the outflow may be colliding with the narrow line region (NLR) gas and exciting it by shocks. In order to evaluate this hypothesis, which is very important for the study of the feedback of the AGN in the circumnuclear region, we compared the observed [N~\textsc{ii}]$\lambda$6583/H$\alpha$, [O~\textsc{iii}]$\lambda$5007/h$\beta$, [O~\textsc{i}]$\lambda$6300/H$\alpha$ and [S~\textsc{ii}]($\lambda$6717+$\lambda$6731)/H$\alpha$ emission-line ratios in the region of the detected outflow with those in the Mappings III shock model library \citep{all08}\footnote{\url{http://cdsweb.u-strasbg.fr/~allen/mappings\_page1.html}}. However, considering the electron density values obtained from the [S~\textsc{ii}]$\lambda$6717/[S~\textsc{ii}]$\lambda$6731 ratio and the gas velocity dispersion values, the observed emission-line ratios cannot be reproduced by the shock heating model.

Since possible shocks from the outflow do not result in the observed emission-line ratios, one could assume that the LINER-like emission-line spectra in the region corresponding to the ionization cone are only due to the photoionization by the central AGN. On the other hand, it is well known that photoionization by post-AGB stars can also generate emission-line spectra typical of LINERs \citep{sta08,era10,cid11}. The observed values of the [N~\textsc{ii}]$\lambda$6583/H$\alpha$, [O~\textsc{iii}]$\lambda$5007/h$\beta$, [O~\textsc{i}]$\lambda$6300/H$\alpha$ and [S~\textsc{ii}]($\lambda$6717+$\lambda$6731)/H$\alpha$ emission line ratios along the ionization cone are in the ranges of 1.7 -- 2.0, 1.5 -- 3.0, 0.25 -- 0.35 and 1.6 -- 2.0, respectively. In order to evaluate whether or not the ionizing radiation from the AGN can explain the observed emission-line spectra along the ionization cone, we performed a simple simulation with version 13.03 of the Cloudy software, last described by \citet{fer13}. We assumed, as an ionizing source, a power law (representing the AGN featureless continuum) with a spectral index of 1.5. The distance between the AGN and the line-emitting region was taken as 70 pc (which is equivalent to the projected distance between these regions) and the electron density of the emitting region was taken as 500 cm$^{-3}$, as obtained from the [S~\textsc{ii}]$\lambda$6717/[S~\textsc{ii}]$\lambda$6731 emission-line ratio. We also assumed a filling factor of 0.01. We then varied the values of the bolometric luminosity of the central source, of the gas metallicity and of the low cut-off energy of the incident radiation on the emitting gas. This last parameter is necessary, in order to account for the extinction effect on the ionizing radiation, before it reaches the emitting gas. At the end, considering a bolometric luminosity of $L_{bol}$ = $2.5 \times 10^{41}$erg s$^{-1}$, a metallicity of z = 0.05 ($Z_{\sun}$ = 0.02 being the solar metallicity) and a low cut-off energy of 40 eV, the resulting [N~\textsc{ii}]$\lambda$6583/H$\alpha$, [O~\textsc{iii}]$\lambda$5007/H$\beta$, [O~\textsc{i}]$\lambda$6300/H$\alpha$ and [S~\textsc{ii}]($\lambda$6717+$\lambda$6731)/H$\alpha$ emission line ratios were 1.9, 1.5, 0.35 and 1.6, respectively, all consistent with the observed values. Therefore, we conclude that the ionizing radiation from the AGN, without taking into account the emission from post-AGB stars, is enough to reproduce the observed emission-line spectra along the ionization cone of the AGN.

The \textit{HST} image in the F658N filter revealed a structure with a morphology of a ring. The PA of the major axis of the ring (44$\degr$ $\pm$ 4\degr) is not compatible with the PA of the major axis of the galaxy disc ($\sim$ 12\degr\!\!.3). In addition, the inclination of the ring, assuming it is circular (66\degr\!\!.2 $\pm$ 2\degr\!\!.3), is not compatible with the inclination of the galaxy disc ($\sim$ 50\degr\!\!.3 -- Hyperleda), which suggests that this ring is not co-planar with the galaxy disc.

The stellar archaeology, performed with spectral synthesis, shows that there is not much difference between the nuclear and circumnuclear regions of NGC 2442 (within the GMOS FOV), both of them showing a predominance of old stellar populations (10 Gyr) with high metallicity (z= 0.02 or 0.05). This absence of recent star formation may be a consequence of the AGN feedback, which, as well established in the literature, can shut off star formation \citep{fab12}. In this case, the AGN feedback is possibly associated with the detected outflow in the GMOS/IFU data cube. The absence of recent star formation also suggests that the bar does not play a significant influence in the star formation in this region. It is worth mentioning that the non detection of young stellar populations could be a problem of degeneracy of the method of the spectral synthesis; therefore, as already mentioned before, this result must be taken with caution.

Since NGC 2442 is a Milky Way morphological twin, it is interesting to compare its central region and the central region of the Galaxy itself. One of the main differences that can be noted is the fact that, unlike NGC 2442, the Galaxy does not harbour an AGN. In addition, the Galaxy shows a fraction of young stars that are found in the nuclear stellar disc, that coincides with the central molecular zone, including O/B stars in the  stellar clusters Arches, Quintuplet, and the nuclear stellar cluster (NSC; \citealt{barbuy}) and Wolf--Rayet stars (e.g. \citealt{gen96,pau01,tan06,rate2020}), which were not detected in NGC 2442. Such kind of young stars are not detected in NGC 2442. On the other hand, the central regions of both the Galaxy and NGC 2442 show a predominance of old stellar populations \citep{gen10}. Another similarity is the presence of circumnuclear structures, such as a nuclear spiral (e.g. \citealt{lo83}). This type of comparison between our Galaxy and the Milky Way morphological twins can contribute  to a better understanding of the formation and evolution of the Galaxy. However, larger samples of Milky Way morphological twins are needed, in order to obtain statistically relevant results.

\section{Conclusions} \label{sec_conclusions}

By gathering information of optical (from the GMOS data cube and \textit{HST} images) and X-ray emission (from the \textit{Chandra} data cube and \textit{XMM--Newton} and \textit{NuSTAR} spectra), we draw an interesting scenario for the nuclear region of the NGC 2442 galaxy, with the main focus of searching for evidence of the interaction between the AGN and the circumnuclear regions. Our main findings are as follows: 

$\bullet$ There is an AGN with an emission-line spectrum compatible with that of LINERs in NGC 2442 nucleus, confirmed by the compact emission of hard and soft X-rays. 

$\bullet$ The analysis of the X-ray data indicates that the AGN is Compton-thick. Most of this obscuration is probably caused by a nearly edge-on torus around the AGN, which, according to the unified model, also explains the apparent absence of broad components in the permitted optical emission lines. This fact together with the optical analysis indicates a type 2 low-luminosity AGN.

$\bullet$ The F658N filter image (H$\alpha$) from \textit{HST} shows a structure that is probably associated with the walls of an ionization cone of the AGN, since its morphology resembles an hourglass. This hypothesis is sustained by the fact that this emission has low velocities. The bipolar morphology of the gas kinematics, together with the high gas velocity dispersion values in this region, suggest the presence of an outflow, which supports the scenario with an ionization cone located in this area.

$\bullet$ The emission-line spectra in the region corresponding to the detected ionization cone are characteristic of LINERs. Shock heating models cannot reproduce the observed emission-line ratios. Therefore, the outflow in this region is not directly associated with the emission-line spectra. On the other hand, a simple photoionization model by a central AGN, assuming a bolometric luminosity of $L_{bol}$ = $2.5 \times 10^{41}$ erg s$^{-1}$, successfully reproduces the observed emission-line ratios.

$\bullet$ Only one ionization cone was detected because the other one (at the opposite side) is highly obscured. In this case, the obscuring material resembles a disc structure, possibly aligned with the inner edge-on torus. This dusty disc structure is probably associated with the nuclear spiral observed in previous studies and represents a feeding mechanism of the AGN.

$\bullet$ A ring was detected in the F658N filter image from \textit{HST} in previous studies. This ring is presumably composed of HII regions and has an elliptical morphology with major axis radius equal to 8.7~$\pm$~0.5 arcsec (979 $\pm$ 80 pc), minor axis radius of 3.51~$\pm$~0.24 arcsec (371 $\pm$ 35 pc) and PA of the major axis equal to 44$\degr$ $\pm$ 4\degr. If the ring is circular, its radius would be equal to 8.7~$\pm$~0.5 arcsec (979 $\pm$ 80 pc) with an inclination of 66\degr\!\!.2 $\pm$ 2\degr\!\!.3.

$\bullet$ The spectral synthesis results indicate the presence of mostly old stellar populations ($\sim$ 10 Gyr), with high metallicity (z = 0.02 and 0.05), and an absence of recent star formation. We propose that this is a consequence of the AGN feedback, associated with the observed outflow, which, as well established in the literature, can shut off star formation.

$\bullet$ Although NGC 2442 is a Milky Way morphological twin, its central region shows many differences relative to the Galaxy, such as the presence of an AGN and the absence of young stellar populations (Wolf--Rayet stars were detected in the central region of the Galaxy). On the other hand, there are also certain similarities, such as the predominance of old stellar populations and the presence of circumnuclear structures. Larger samples of Milky Way morphological twins will allow more detailed comparisons and provide information about the formation and evolution of this type of galaxies.

\section{Data availability}

The data used in this paper are available in public archives: the GMOS data are from Gemini Observatory Archive at \url{https://archive.gemini.edu/searchform}\footnote{Developed by Paul Hirst, Ricardo Cardenes, Adam Paul, Petra Clementson and Oliver Oberdorf.}, and the raw and treated data cubes from GMOS used in this work can be found at \url{https://syr.us/BMb}. The \textit{HST} data are from Mikulski Archive from Space Telescopes at \url{https://archive.stsci.edu/hst/}, the \textit{Chandra} data are from  Chandra Data Archive at \url{https://cda.harvard.edu/chaser/}, the \textit{NuSTAR} data are from HEASARC at \url{https://heasarc.gsfc.nasa.gov/docs/nustar/nustar_archive.html} and the \textit{XMM--Newton} data are from XMM--Newton Science Archive at \url{http://nxsa.esac.esa.int/nxsa-web/#home}.

\section*{Acknowledgements}

This work is based on observations obtained at the Gemini Observatory (processed using the Gemini \textsc{IRAF} package), which is operated by the Association of Universities for Research in Astronomy, Inc., under a cooperative agreement with the NSF on behalf of the Gemini partnership: the National Science Foundation (United States), the National Research Council (Canada), Comisión Nacional de Investigación Científica y Tecnológica (Chile), the Australian Research Council (Australia), Minist\'erio da Ci\^encia, Tecnologia e Inova\c{c}\~ao (Brazil), Ministerio de Ciencia, Tecnolog\'ia e Innovaci\'on Productiva (Argentina), and Korea Astronomy and Space Science Institute (Republic of Korea). This work is based on observations made with the NASA/ESA Hubble Space Telescope, obtained from the Data Archive at the Space Telescope Science Institute, which is operated by the Association of Universities for Research in Astronomy, Inc., under NASA contract NAS 5-26555. This research has also made use of the NASA/IPAC Extragalactic Database (NED), which is operated by the Jet Propulsion Laboratory, California Institute of Technology, under contract with the National Aeronautics and Space Administration. This research has made use of data obtained from the Chandra Data Archive of observations made by the Chandra X-ray Observatory. This research has made use of NASA's Astrophysics Data System. This involved observations obtained with \textit{XMM--Newton}, an ESA science mission with instruments and contributions directly funded by ESA Member States and NASA. This research has made use of data and/or software provided by the High Energy Astrophysics Science Archive Research Center (HEASARC), which is a service of the Astrophysics Science Division at NASA/GSFC. We thank CNPq (Conselho Nacional de Desenvolvimento Cient\'ifico e Tecnol\'ogico), under grant 141766/2016-6, and FAPESP (Funda\c{c}\~ao de Amparo \`a Pesquisa do Estado de S\~ao Paulo), under grants 2011/51680-6 and 2020/13315-3, for supporting this work. YD acknowledges the financial support from Programa de Investigación Asociativa (PIA) of the Comisión Nacional de Investigación Científica y Tecnológica (CONICYT) ACT172033.  ELN acknowledges financial support from the National Agency for Research and Development (ANID) Scholarship Program DOCTORADO NACIONAL 2020 - 21200718. We thank Prof. Beatriz Barbuy and Dr. Daniel May for the kindness of reviewing this work. This work is dedicated to the memory of Prof. Jo\~ao Steiner, who passed away in 2020 September 10.



\bibliographystyle{mnras}
\bibliography{references} 


\appendix

\section{Maps obtained with the \textsc{starlight} software applied to the GMOS data cube}\label{mapas_starlight}

\begin{figure}
\begin{center}

  \includegraphics[scale=0.36]{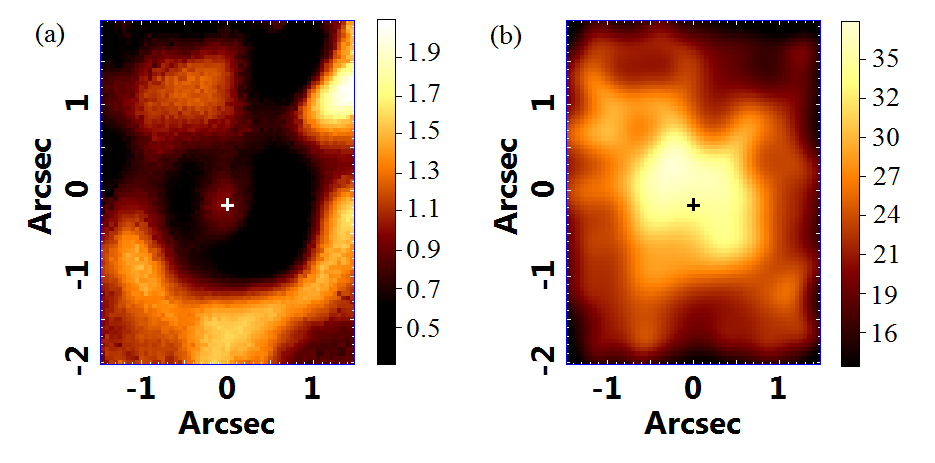}
  \caption{Maps obtained from the spectral synthesis using the \textsc{starlight} software on the GMOS data cube of (panel a) extinction and (panel b) S/N. The crosses indicate the position of the centre of [O~\textsc{i}]$\lambda$6300 emission, and its size represents the 3$\sigma$ uncertainty. \label{NGC2442_maps_starlight2}}
  
\end{center}
\end{figure}

As described in sections \ref{sec_starlight} and \ref{sec_starlightresults}, we applied the spectral synthesis to the GMOS data cube, in order to study the stellar archaeology of the central region of NGC 2442. One of the results of this method is the extinction map, which is one of the parameters that the spectral synthesis fits in order to delineate the observed stellar continuum. Fig.~\ref{NGC2442_maps_starlight2}(a) shows this map, in which we can see a structure similar to a ring/spiral of extinction in the central region of NGC 2442. The values are quite high and show, saving the uncertainties at the edges of the FOV, such that the southern part has higher extinction when compared to the northern part. This result is compatible with what we observe in the \textit{HST} images: a redder emission towards south. 

The map of S/N (Fig.~\ref{NGC2442_maps_starlight2}b) shows that the whole GMOS FOV has a good S/N (higher than 10) and the values are higher in the central portion (up to 40).

\section{X-ray spectra modelling details}\label{modelsxray}

Here we describe with more details the analysis of \textit{XMM--Newton} and \emph{NuSTAR} spectra, as said previously, we followed the method used by  \citet{dia20}. First, we started our X-ray spectral analysis with an absorbed power law and absorbed \texttt{pexmon} in the \textsc{xspec} program: \texttt{constant*phabs*[cabs*zphabs(soft)*powerlaw + cabs*zphabs(hard)*pexmon]}. In this model, the \texttt{phabs} component is associated with absorption from our Galaxy and it was fixed to the predicted value using the N$_{\rm H}$ tool within \textsc{ftools} \citep{dic90}. We included a multiplicative constant normalisation between \emph{NuSTAR} FPMA-FPMB and \emph{XMM–Newton} EPIC-PN to account for calibration uncertainties between the instruments. The component called \texttt{zphabs(soft)} is related to the absorption on the extended scattered power law and \texttt{zphabs(hard)} is associated with the absorption on the nuclear region and \texttt{cabs} is the absorption related with a optically-thin Compton scattering. A similar model was used by \citet{gonzalezmartin09}, showing that it is a good representation of the X-ray spectra of 82 LINERs obtained with \textit{XMM--Newton} and \textit{Chandra}. \texttt{pexmon} \citep{nandra2007} represents both the reflected and intrinsic emission and the scattered power law has a normalisation of a few percent of the primary power law with an identical slope. This fit results in $\chi^{2}$=261.6 for 215 degrees of freedom (d.o.f.). This model fails to adequately fit the spectral continuum, leaving obvious structured residual at energies below 10.0 keV. We added a \texttt{MEKAL} component (a thermal component at soft energies) to the model to check whether this component improves the fit. We found a $\chi^{2}$=211.1 for 212 d.o.f.,  $\Gamma$<2.41, E$_{cut}$<30 keV and reflection fraction R$_{f}$=1.46$_{-1.31}^{+2.48}$. Also, an F-test was used to confirm an improvement of the fit. We found an F-test probability of 1.34$\times$10$^{-10}$. Thus, the best-fitting model is the absorbed \texttt{power law + MEKAL} model that represents the data at energies below 10.0 keV.

 In order to fit the data with a more representative physical model, we studied reflection models which might come from a neutral reflector as modelled by \textit{Borus} \citep{balokovic18}. This is a radiative transfer code that calculates the reprocessed continuum of photons that are propagated through a neutral and cold medium. In this work, we used the geometry that corresponds to a smooth spherical distribution of neutral gas, with conical cavities along the polar directions, also called, \texttt{Borus02}. The reflected spectrum of the reflector is calculated for a cut-off power law illuminating continuum, where the E$_{\rm{cut}}$, $\Gamma$ and the normalisation are free parameters. The opening angle of the cavities ($\theta_{\rm{tor}}$), as well as the column density ($\log(\rm{N_{H,tor}})$) and the inclination of the torus $\theta_{\rm{incl}}$, are also free parameters. In this case, we tied $\theta_{\rm{tor}}$ to $\theta_{\rm{incl}}$ to ensure a direct view of the central engine. We modelled the coronal emission separately with a cut-off power law under a neutral absorber with \texttt{cabs*zphabs}. The baseline model used to fit this source was:

\begin{equation}
   C*N_{Gal}*(MEKAL + N_{H,S}*PL + Borus + N_{H,H}*cPL),
\end{equation}

where C represents the cross-calibration constant between the instruments and N$_{Gal}$ is the Galactic absorption (\texttt{phabs} in \rm{XSpec}); N$_{H,S}$ is the column density of absorbing material acting on the scattered power law, PL is the power law of the scattered component; N$_{H,H}$ is the absorbing material that acts on the nuclear components (power law and torus); and cPL is a cut-off power law (\texttt{cutoffpl} in \textsc{xspec}) representing the primary X-ray emission and \textit{Borus} represents the reflection model.

\bsp	
\label{lastpage}
\end{document}